\documentclass[aps,prl,twocolumn,groupedaddress,amsmath,amssymb]{revtex4-1}
\usepackage{graphicx}
\usepackage{dcolumn}
\usepackage{bm}
\usepackage{color}
\usepackage{dsfont}

\newcommand{\+}{\uparrow}
\renewcommand{\-}{\downarrow}
\renewcommand{\>}{\rangle}
\newcommand{\<}{\langle}
\newcommand{\ahatdag}{\hat{a}^{\dag}}
\newcommand{\ahat}{\hat{a}}
\newcommand{\bhatdag}{\hat{b}^{\dag}}
\newcommand{\bhat}{\hat{b}}
\newcommand{\chatdag}{\hat{c}^{\dag}}
\newcommand{\chat}{\hat{c}}

\newcommand{\nhat}{\hat{n}}

\newcommand{\zhat}{\hat{z}}
\newcommand{\Id}{\mathds{1}}

\newcommand{\Hh}{\mathcal{H}}
\newcommand{\Jj}{\mathcal{J}}

\newcommand{\Ss}{\mathcal{S}}

\begin{document}


\title{A many-body interpretation of Majorana bound states, and conditions for their localisation.}


\author{Thomas E. O'Brien}
\email{obrien@lorentz.leidenuniv.nl}
\affiliation{Instituut-Lorentz for Theoretical Physics, Leiden University, Leiden, The Netherlands}%
\affiliation{The Perimeter Institute for Theoretical Physics, Waterloo, Canada}%
\affiliation{School of Mathematics and Physics, University of Queensland, Brisbane, Queensland, Australia}%
\author{Anthony R. Wright}%
\affiliation{School of Mathematics and Physics, University of Queensland, Brisbane, Queensland, Australia}%

\date{\today}

\begin{abstract}
We derive a condition for the existence of completely or exponentially localised Majorana bound states (with the potential for non-Abelian statistics) in a generic many-body system. We discuss the relationship between the existence of these operators and the protection of the ground state degeneracy from local perturbations. We use our methods to study the exponential decay of the Majorana bound states in the non-interacting Kitaev chain, finding complete agreement between our many-body calculation and single-particle results. We then apply these results to various interacting systems which have previous evidence for Majorana bound states.
\end{abstract}


\maketitle


In 1936 a real solution to the Dirac equation was discovered by Ettore Majorana \cite{ref:Majorana}. This corresponds to a class of fermions that are their own anti-particles. Though physical examples of these Majorana fermions have eluded us to this date, the zero-energy Majorana bound state (MBS) has been proposed as a condensed matter quasiparticle analogue. This has generated a large amount of interest due to the suggestion it may obey non-Abelian statistics \cite{ref:Ivanov,ref:ReadGreen}, with implications for both fundamental physics and quantum computation \cite{ref:SimonReview}. Over the last few years, the set of hypothetical systems supporting Majorana bound states has grown in leaps and bounds \cite{ref:AliceaReview,ref:SauDasSarma,ref:Alicea,ref:KitaevHoneycomb}. Experiments have demonstrated the existence of a zero-energy state in superconducting wires \cite{ref:Mourik,ref:Das} and atomic chains \cite{ref:Princeton}, and schemes have been developed to implement and observe braiding experimentally \cite{ref:BVH,ref:Hyart}.

In the last few years, interest has developed in interacting systems that support MBSs \cite{ref:BeenakkerWire,ref:AliceaWire,ref:LB,ref:DoubleWirePaper,ref:TonyMennoPaper,ref:DQDPaper,ref:KST,ref:CCS}. Evidence for topologically nontrivial excitations has been found in Josephson currents \cite{ref:BeenakkerWire,ref:TonyMennoPaper}, entanglement spectra and correlation functions \cite{ref:DoubleWirePaper,ref:LB,ref:CCS}, or by continuously tuning to a non-trivial topological phase in a non-interacting limit\cite{ref:DQDPaper,ref:KST}. Importantly, in some of these systems, strong interactions are \emph{required} for the appearance of these effects. However, in the single-particle formalism, data about topological protection and the potential for non-trivial braiding may be read from the solutions to the single-particle Hamiltonian. In the many-body case, this data is obscured.

In this work, we present a method for determining this data for a generic many-body system. We derive conditions for the existence of an operator that we claim corresponds to a localised MBS. We demonstrate how the form of this operator may be derived when these conditions are satisfied, and discuss the implications of the existence of this operator on ground state protection. We then apply these methods to the previous studied models of \cite{ref:DQDPaper,ref:TonyMennoPaper,ref:DoubleWirePaper,ref:KST,ref:CCS}.

For the purposes of this paper, we will concern ourselves with fermionic systems with discrete sets of sites, and assume all Hamiltonians conserve fermion number parity. Then, for a zero-energy Majorana bound state, we require a degeneracy between the lowest even particle number ($|\Psi_E\>$) and odd particle number ($|\Psi_O\>$) many-body states of our system. To simplify the discussion, we will assume no additional degeneracies exist; though these are necessary for non-Abelian statistics, our methods immediately generalise.
Given such a system with Hamiltonian $\Hh$, suppose we wish $\gamma$ to be an operator corresponding to the creation or destruction of an MBS. We could require $\gamma$ to be Hermitian and unitary, to excite $|\Psi_E\>$ to $e^{i\phi}|\Psi_O\>$, and to commute with $\Hh$ \cite{ref:SimonReview,ref:Kitaev}. These conditions are satisfied by the MBS in non-interacting systems such as \cite{ref:Kitaev,ref:ReadGreen,ref:Ivanov}. 
However, these conditions assume that excitations in the system behave independently of the state of the system, which is not necessarily the case when interactions are present. This can be seen in the Hubbard model in the strong-coupling limit: at half-filling the system is a Mott insulator, but the addition or removal of a single electron turns the system into a conductor, with a dramatically different energy spectrum.
As such, we restrict our conditions to applying only on the subspace spanned by our degenerate ground states. This is equivalent to assuming that the gap between these states and those at higher energy is much greater than the temperature. The above conditions then become
\begin{align}
\gamma^{\dag}=\gamma,\label{eqn:RMajoranaCondition1}\\
\<\Psi_A|\gamma^2|\Psi_B\>&=\delta_{AB},\label{eqn:RMajoranaCondition2}\\
\<\Psi_A|\gamma|\Psi_B\>&=(1-\delta_{AB})e^{\pm i\phi},\label{eqn:RMajoranaCondition3}\\
\<\Psi_A|\left[\Hh,\gamma\right]|\Psi_B\>&=0,\label{eqn:RMajoranaCondition4}
\end{align}
where $A$ and $B$ are either $O$ or $E$, and the sign on the phase shift is dependent on whether we are going from $|\Psi_E\>$ to $|\Psi_O\>$ or vice-versa.

These conditions can be simplified somewhat. Equation \ref{eqn:RMajoranaCondition4} is a consequence of equation \ref{eqn:RMajoranaCondition3} and the ground states being degenerate. Then, equation \ref{eqn:RMajoranaCondition3} implies that we can write $\gamma|\Psi_O\>=e^{\pm i\phi}|\Psi_E\>+|\xi\>$ with $\<\Psi_E|\xi\>=0$, and expanding equation \ref{eqn:RMajoranaCondition2} with $A=B=O$ gives that $|\xi\>=0$. That is, $\gamma$ is still required to strictly excite between $|\Psi_E\>$ and $|\Psi_O\>$. This also makes physical sense, as otherwise $\gamma$ would not  be an excitation between our ground states. Finally, with this new restriction equation \ref{eqn:RMajoranaCondition2} is automatically fulfilled. We then have a set of \emph{reduced Majorana conditions}
\begin{align}
\gamma^{\dag}&=\gamma,\label{eqn:MajoranaCondition1}\\
\gamma|\Psi_A\>&=e^{\pm i\phi}|\Psi_{\bar{A}}\>,\label{eqn:MajoranaCondition2}
\end{align}
with $\bar{A}$ being the opposite parity to $A$. 

If $\gamma$ satisfies these, we say it is a \emph{Majorana operator}. However, if every Majorana operator corresponded to an MBS, then an MBS could be found whenever there exists a ground state degeneracy, for the following Majorana operator always satisfies the reduced Majorana conditions:
\begin{align}
\gamma_{\phi}&=\cos(\phi)\left(|\Psi_O\>\<\Psi_E|+|\Psi_E\>\<\Psi_O|\right)\nonumber\\&+i\sin(\phi)\left(|\Psi_O\>\<\Psi_E|-|\Psi_E\>\<\Psi_O|\right).
\label{eqn:BareMajorana}
\end{align}
Indeed, any operator of the form $\gamma_{\phi}+G$, with $G$ a self-adjoint operator that does not act on our ground states, will be a Majorana operator (and every Majorana operator in our system is of this form). We call $\gamma_{\phi}$ the \emph{bare Majorana}.

This result by itself is thoroughly uninteresting, and is definitely not unknown. Clearly the conditions currently stated are not sufficient to guarantee the system can at all be described as supporting MBSs with exotic properties. As such, we turn to these properties for further restrictions. We consider the protection of the ground state degeneracy, and the possibility for braiding statistics.

The notion of braiding of excitations is intrinsically tied with locality. Given two excitations, we wish to define a non-trivial closed path that these excitations can travel in real space such that the system obtains a non-trivial geometric phase upon completion. This cannot be done unless the location of the particles is well-defined. But, if we attach the geometric phase to a creation operator - $\chatdag\rightarrow e^{i\alpha}\chatdag$, then the corresponding number operator $\hat{n}=\chatdag\chat$ remains invariant. This implies that if our Majorana operator contains products of number operators with single creation and annihilation operators, only the latter must be localised to allow braiding.

Let us fix a region of our system $R$ that we wish to localise our Majorana operator $\gamma$ within, and let $\Jj$ be the set of sites outside $R$. We require that if $j$ is a site in $\Jj$, $\gamma$ contains no single products of $\chatdag_j$ or $\chat_j$. This is equivalent to requiring that $\gamma$ commutes with $\hat{n}_j$. Then $\gamma$ will commute with every product $\hat{n}_{J}=\prod_{j\in J}\hat{n}_j$ of number operators from subsets $J\subset\Jj$. Any matrix element of the commutator $\left[\hat{n}_J,\gamma\right]$ must then be zero, and we calculate
\begin{align}
\<\Psi_O|\left[\hat{n}_{J},\gamma\right]|\Psi_E\>&=0\nonumber\\\<\Psi_O|\hat{n}_{J}\gamma|\Psi_E\>&=\<\Psi_O|\gamma\hat{n}_{J}|\Psi_E\>\nonumber\\
|\<\Psi_O|\hat{n}_{J}|\Psi_O\>&-\<\Psi_E|\hat{n}_{J}|\Psi_E\>|:=d_J=0.\label{eqn:LocalisationCondition}
\end{align}
We call $d_J$ the \emph{number difference on $J$}, and the requirement for $d_J=0$ for all $J\subseteq\Jj$ the \emph{localisation condition} for our system to admit a Majorana operator $\gamma$ localised \emph{away} from $\Jj$ (or \emph{to} $R$). Amazingly, although we have currently only proven that this condition is \emph{necessary} for a Majorana operator, it is sufficient also. In the supplimentary material, we prove this by giving a method of constructing the Majorana operator, including explicit formulae for small systems.

Let us now consider the effect of a perturbation $V$ upon the Hamiltonian. We assume that $V$ continues to conserve Fermion number parity, as breaking this will cause quasiparticle poisoning. Then, $V$ cannot couple our two ground states (as they are of different parity), and so to first order in perturbation theory, the induced gap between the ground states is
\begin{equation}
\Delta=|\<\Psi_E|V|\Psi_E\>-\<\Psi_O|V|\Psi_O\>|.
\label{eq:ProtCond}
\end{equation}
This shows that the Majorana condition on $J$ is equivalent to the requirement that our system is protected against small local perturbations of the form $V=\delta n_J$. 

This is not the only type of local perturbation that can occur in our system. Thus, the presence of localised Majorana operators as defined do not entirely protect the system. To get more insight into this statement, we can insert a Majorana operator $\gamma$ into equation \ref{eq:ProtCond}:
\begin{equation}
\Delta=|\<\Psi_O|\left[\gamma V\gamma - V\right]|\Psi_O\>|.
\end{equation}
We see that an equivalent condition for the protection of the degeneracy is the existence of some (not necessarily localised) Majorana operator $\gamma$ that commutes with $V$. In a system where the Majorana condition holds, but the ground state is not protected against \emph{all} local perturbations, an experiment to observe the effect of the MBSs in the system will need to be isolated from these unprotected perturbations.

The above discussion implies that a non-trivial system must be able to localise Majorana operators away from \emph{all} local regions (i.e. $d_J=0$ for all local $J$). This makes sense, as our degeneracy supports two anti-commuting Majorana operators, constructed from the bare Majoranas $\gamma_0$ and $\gamma_{\pi/2}$ (note that it is impossible with only one degeneracy to find more than two mutually anti-commuting Majorana operators). When these operators can be spatially separated, any local perturbation that commutes with either one or the other cannot lift the degeneracy without first pushing the operators together.

\begin{figure}
\includegraphics[width=\columnwidth]{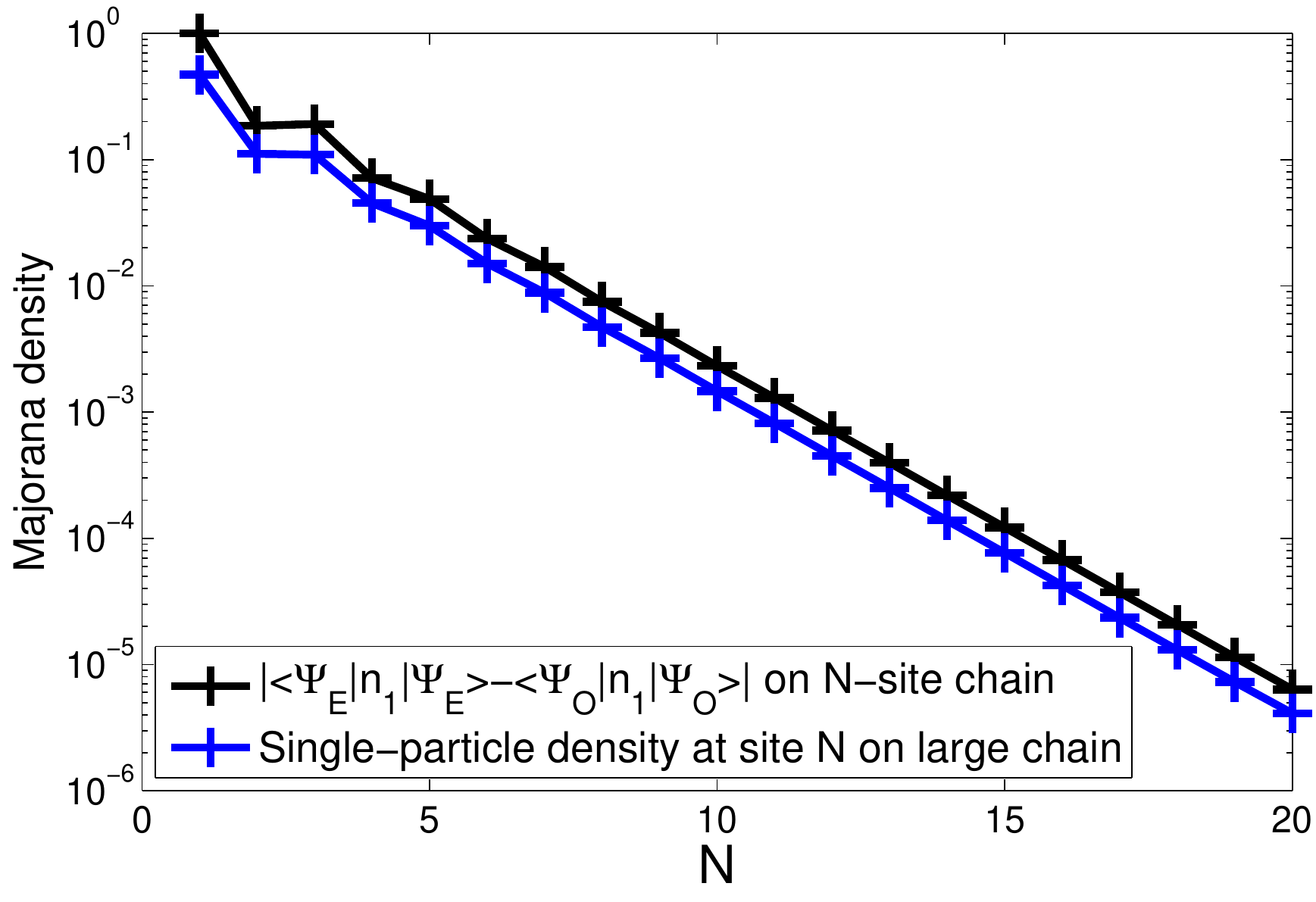}
\caption{Measurement of the exponential decay of the Majoranas in the Kitaev chain via the ground state number difference $d_1$ (black line), and via the solution to the single-particle Hamiltonian (blue line). Although the single-particle data only exists in the non-interacting regime, the $d_1$ scaling is measurable in any scalable many-body system.}
\label{fig:KitChain}
\end{figure}

Now, let us consider a non-local region $R$ in our system such that each subregion $S\subset R$ satisfies $d_S=0$, but $d_R\neq 0$. Then, we can consider $R$ to be a 'Majorana-supporting subsystem'; every Majorana we write must have some density in $R$. Note that such an $R$ must exist, although it may be the entire system. When $R$ is not the entire system, there is the possibility for Majorana operators to either take the forms $\gamma_R$ (if there is no number difference outside $R$), $\gamma_{R}+\gamma_{\bar{R}}$ or $\gamma_{R}\gamma_{\bar{R}}$ (or some combination thereof) where $\bar{R}$ is the region outside $R$. In the first situation, the MBS can be considered to be split into parts, whilst in the second situation, the MBS will be an effective description of the system inside $R$.

Within each region $R$ however, it is not at all obvious where the MBS itself lies; we may remove all but \emph{any} one site within the region. As an example, consider the Kitaev chain with Hamiltonian
\begin{equation}
\Hh_{Kitaev}=-\sum_{j=1}^{N-1}(t\chatdag_{j}\chat_{j+1}+\Delta\chatdag_j\chatdag_{j+1}+h.c.)+\mu\sum_{j=1}^N\hat{n}_j.
\label{eqn:KitaevChain}
\end{equation}
in the special point where $\Delta=t$ and $\mu=0$, the following Majoranas are localised to every site $1\leq j\leq N$:
\begin{align*}
\gamma_{j,\phi}&=\cos(\phi)(\chatdag_j+\chat_j)\prod_{l=1}^{j-1}(1-2\hat{n}_l)\\
&+i\sin(\phi)(\chatdag_{j}-\chat_j)\prod_{l=j+1}^N(1-2\hat{n}_l).
\end{align*}
This is clearly a problem with our description. To rectify it, we consider the experimental probes that might be used to determine the MBS position. These usually take the form of some transport measurement; we connect the system to a lead which breaks local parity conservation, and attempt to locally excite between the two ground states. Then, we are looking for a \emph{probe density} $\rho_r=\<\Psi_E|\hat{p}_{r}|\Psi_O\>\neq 0$, with $\hat{p}_r$ an operator entirely confined to a small region $r$ that depends on the system and probe in question, but with no restriction to satisfy the Majorana conditions. We expect to see exponentially localised $\rho_r$, as it is in the Kitaev chain with $\hat{p}_r=\chatdag_i$ for $r=\{i\}$.

It may well be asked here why this is not a sufficient condition in the first place. However, we stress that we can only find \emph{Majorana-like} excitations when the localisation condition is satisfied for all local regions. Thus, a full study should include demonstrating that all Majorana-supporting subsystems are non-local, and then investigating both degeneracy protection and the Majorana location.

In a realistic system, we expect an MBS to have exponentially decaying tails, and so all MBSs will exist everywhere (and braiding will have an exponentially small chance of errors). This should cause a breaking of the localisation condition in local areas, but only by an amount that is exponentially small in the system size, as this measures the minimum density of \emph{all} MBSs on a given site. To check this, in figure \ref{fig:KitChain} we plot the deviation $d_1=|\<\Psi_E|\hat{n}_1|\Psi_E\>-\<\Psi_O|\hat{n}_1|\Psi_O\>|$ for the Kitaev chain as we increase the system size $N$, in the topologically non-trivial phase (Hamiltonian given in equation \ref{eqn:KitaevChain}, with $t=0.7\Delta$, $\mu=0.1\Delta$). Comparing this to the density of the single particle Majorana operators in a larger chain ($50$ sites) finds complete agreement up to a constant factor. As this procedure can be repeated for a \emph{many-body} system of adjustable size, we have a method of measuring the exponential decay of Majoranas in a realistic system. However, as noted before, this procedure by itself will not locate individual MBS within the system.

We now apply our procedure to the interacting Kitaev chain studied in \cite{ref:KST,ref:CCS}. Taking the formalism of \cite{ref:KST}, the Hamiltonian of the system is
\begin{equation}
\Hh=\Hh_{Kitaev}+U\sum_{j=1}^{L-1}(2\hat{n}_j-1)(2\hat{n}_{j+1}-1),
\end{equation}
where $L$ is the length of the chain. A ground state degeneracy is obtained when the bulk chemical potential is $\mu=\mu^*:=4\left(U^2+tU+(t^2-\Delta^2)/4\right)^{1/2}$, and the edge sites have a chemical potential half this size \cite{ref:KST}. At this point, we can calculate the number difference and the probe density, with probe $\hat{p}_n=\chatdag_n$:
\begin{align}
d_i&=2\frac{\alpha^2(1-\alpha^2)^{L-1}}{(1+\alpha^2)^L},\\
\rho_n&=\alpha\left[\frac{(1-\alpha^2)^{n-1}}{(1+\alpha^2)^n}-\frac{(1-\alpha^2)^{L-n}}{(1+\alpha^2)^{L-n+1}}\right].
\end{align}
Here, $\alpha=[\mu^*(\sqrt{(2\Delta/\mu^*)^2+1}+1)/(2\Delta)]^{1/2}$. We see the expected exponential decay of both quantities. 

We now apply our method to the double quantum dot of \cite{ref:DQDPaper,ref:TonyMennoPaper}. The effective Hamiltonian of this system is
\begin{align}
\Hh&=-E_Z\sum_{j=1}^2(\nhat_{j\+}-\nhat_{j\-})+U\sum_{j=1}^2\nhat_{j\+}\nhat_{j\-}\nonumber\\&+t\sum_{\sigma=\pm}\left(\cos\left(\frac{\theta}{2}\right)\chatdag_{1,\sigma}\chat_{2,\sigma}+\sigma\sin\left(\frac{\theta}{2}\right)\chatdag_{1,\sigma}\chat_{2,\bar{\sigma}}+\text{h.c.}\right)\nonumber\\&+\Delta e^{i\frac{\phi_+}{2}}\cos\left(\frac{\phi_-}{2}\right)\left(\sum_{i=1}^2\chatdag_{i\+}\chatdag_{i\-}+\sum_{\sigma=\pm}(\sigma\cos\left(\frac{\theta}{2}\right)\chatdag_{1\sigma}\chatdag_{2\bar{\sigma}}\right.\nonumber\\&\left.-\sin\left(\frac{\theta}{2}\right)\chatdag_{1\sigma}\chatdag_{2\sigma}+\text{h.c.})\right).
\label{EffectiveHamiltonian}
\end{align}
In the infinite $U$ and vanishing $E_Z$ limit, there is a degeneracy when $\Delta=\sqrt{2}t$ \cite{ref:TonyMennoPaper}. The ground states are then given by
\begin{align}
|\Psi_E\>&=\frac{1}{2}\left[\sqrt{2}e^{i\phi_+/4}-e^{-i\phi_+/4}\cos(\theta/2)(\chatdag_{1\+}\chatdag_{2\-}+\chatdag_{2\+}\chatdag_{1\-})\right.\nonumber\\&\left.+e^{-i\phi_+/4}\sin(\theta/2)(\chatdag_{1\+}\chatdag_{2\+}+\chatdag_{1\-}\chatdag_{2\-})\right]|v\>,\\
|\Psi_O\>&=\frac{1}{\sqrt{2}}\left[\cos(\theta/4)(\chatdag_{2\+}-\chatdag_{1\+})\right.\nonumber\\&\left.\hspace{3cm}+\sin(\theta/4)(\chatdag_{2\+}+\chatdag_{1\+})\right]|v\>.
\end{align}
The description of the MBSs in this system is complicated by the addition of spin. If we redefine our spin axes by writing $\chatdag_{j\sigma\rho}=\cos(\rho_j/2)\chatdag_{j\sigma}+\sigma\sin(\rho_j/2)\chatdag_{j\bar{\sigma}}$, the localisation condition is satisfied with $\rho_j=\frac{\pi}{2}+(-1)^j\frac{\theta}{2}$.

That we can only do this is indicative of a lack of protection in the system. Specifically, the system is not protected against on-site magnetic fluctuations save in the $\rho_j$ direction. The system \emph{is} protected against non-magnetic fluctuations of the form $\hat{n}_{j\+}+\hat{n}_{j\-}$. However, as a Zeeman potential $E_Z\approx k_BT$ is required to lift the problematic Kramers degeneracy, a spin-orbit coupling of near $\theta=\pi$ is required so that this does not break the MBS-supporting degeneracy also. At this point, the probe density $\<\Psi_E|\chatdag_{j\sigma}|\Psi_O\>=\frac{1}{\sqrt{2}}$ is independent of $j$ and $\sigma$, and we can write down exact Majorana operators due to the small size of the system (details in the supplementary material). For example, a Majorana operator localised to site $2$ with $\phi=\phi_+=0$ is
\begin{align*}
\gamma&=\zhat_{2\-}(\frac{1}{\sqrt{2}}\zhat_{1\+}\zhat_{1\-}+\hat{n}_{1\+}\zhat_{1\-})(\chatdag_{2\+}+\chat_{2\+})\\&+\zhat_{2\+}(\frac{1}{\sqrt{2}}\zhat_{1\+}\zhat_{1\-}-\hat{n}_{1\-}\zhat_{1\+})(\chatdag_{2\-}+\chat_{2\-}).
\end{align*}

\begin{figure}
\includegraphics[width=\columnwidth]{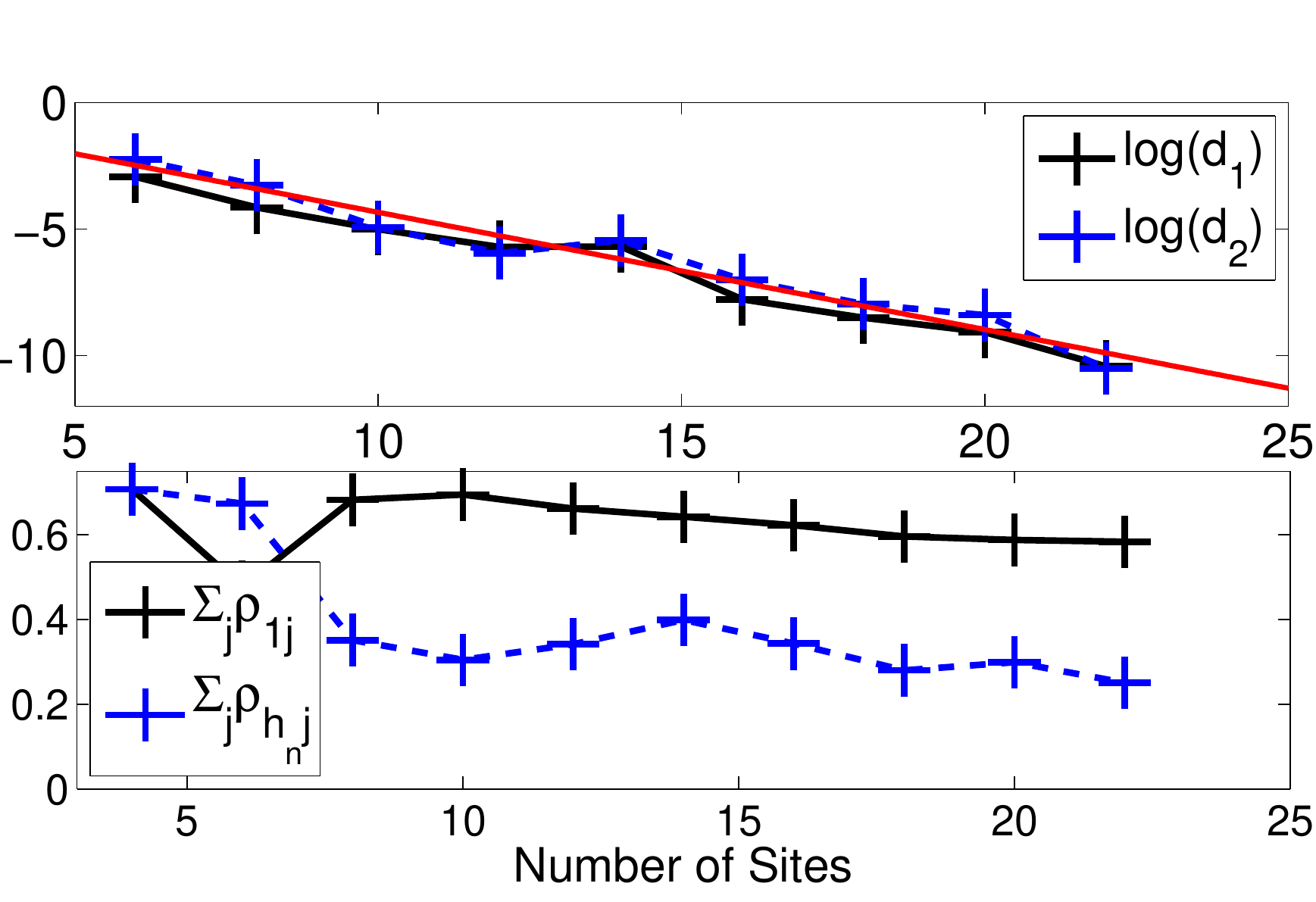}
\caption{(top) Demonstration of exponential decay in the double wire of \cite{ref:DoubleWirePaper} via the number differences $d_1$(black, solid) and $d_2$ (blue, dashed). Red line is an exponential fit to $d_2$ for ease of viewing. (bottom) plot of probe density as a function of system size, both in the middle and at the edges of the sample (probe defined in text).}
\label{fig:DoubleWire}
\end{figure}

Finally, we consider the system of \cite{ref:DoubleWirePaper}. This consists of two spinless wires $a$ and $b$, with creation operators $\ahatdag_i$ and $\bhatdag_i$, and the \emph{number-conserving} Hamiltonian
\begin{align}
\Hh&=-\sum_{j=1}^N\left[t_a\ahatdag_j\ahat_{j+1}-t_b\bhatdag_j\bhat_{j+1}+W\bhatdag_j\bhatdag_{j+1}\ahat_j\ahat_{j+1}+\text{h.c.}\right]
\end{align}
In \cite{ref:DoubleWirePaper}, evidence of a topological phase in this system was found, in the large $N$ limit, at around the $1/3$ filling fraction, and near $t_a+t_b=W$. In this region, the two ground states in the even and odd $a$-wire parity sectors have an energy gap that grows exponentially small in the system size \cite{ref:DoubleWirePaper}. In figure \ref{fig:DoubleWire}(top), we plot exact results of the number difference $d_1$ in this phase ($t_a=t_b=0.5W$, $n\approx 1/3$), as a function of system size $(2N)$. This shows signs of exponential decay, however finite size effects cause non-linearities in the curve.

As the particle number is the same in both of the states we are interested in, our probe cannot be $\ahatdag_i$. Instead, an excitation must involve hopping particles between the $a$ and $b$ wires. Thus, we propose $\hat{p}_{ij}=\ahatdag_i\bhat_j$ as a potential probe of the system. If our Majorana bound states are localised at the edges of the wire, we would expect $\rho_{ij}=|\<\Psi_E|\hat{p}_{ij}|\Psi_O\>|$ to be approximately $0$ unless both $i$ and $j$ are near the edges of the system. Unfortunately, the small system sizes we use here obscure these signatures. In figure \ref{fig:DoubleWire}(bottom), we plot $\sum_j\rho_{1j}$ and $\sum_j\rho_{h_Nj}$, where $h_N$ is taken from the middle of the system with system size $N$ (other parameters are $t_a=t_b=0.5 W$, $n\approx 1/3$). We see that the probe density in the middle of the system is much smaller than at the edges. However we do not have enough data to be sure that one survives whilst the other drops to zero in the large $N$ limit. As DMRG studies have been performed on this system on up to $N=36$ \cite{ref:DoubleWirePaper}, definite results should be obtainable. From the above discussion, we expect the Majorana operators to take the form $\sum\ahatdag_i\bhat_jN+h.c.$ with $i$ and $j$ on the same ends of the wire, and $N$ a product of number operators. 

The results derived in this paper do not present a completed study of an MBS in the many-particle framework. It is possible to immediately include larger numbers of degeneracies; with $2^n$ ground states we will obtain $2n$ MBSs by repeating our procedure on pairs of ground states. However, other important questions remain unanswered, like the exact nature of braiding, fusion and measurement in this formalism. We hope these concepts could be inherited naturally from existing work.
\acknowledgments{T.E.O. wishes to thank J.A. Hutasoit, B. van Heck, B. Tarasinski and C.W.J. Beenakker, for enlightening discussions. We also thank H. Katsura, D. Schuricht and L. Marnham for manuscript feedback. T.E.O. was financially supported by the Perimeter Scholars International program and the University of Leiden. A.R.W. was financially supported by a University of Queensland Postdoctoral Research Fellowship.}
\bibliography{Paper}

\appendix
\section{Proof that localisation condition is exact}\label{app:SuffProof}
Our proof proceeds by giving a method to construct the Majorana operator when the localisation condition is satisfied, and proving that the method works consistently. Let us assume that we have satisfied the localisation condition for $\Jj$. Then, for all $J\subset\Jj$, we can define $n_J:=\<\Psi_E|\hat{n}_J|\Psi_E\>=\<\Psi_O|\hat{n}_J|\Psi_O\>$. Then, let $A=O$ or $E$, and define
\begin{equation}
|\Phi_{A,J}\>:=\frac{1}{\sqrt{n_J-n_J^2}}\left(\hat{n}_J-n_J\right)|\Psi_A\>.
\end{equation}
Then, fix $K,J\subset\Jj$, and we can calculate
\begin{align*}
\hat{n}_K|\Phi_{A,J}\>&=\frac{1}{n_J-n_J^2}\left(\hat{n}_{J\cup K}-n_J\hat{n}_K\right)|\Psi_A\>\\
&=\frac{1}{n_J-n_J^2}\left[(n_{K\cup J}-n_{K\cup J}^2)^{1/2}|\Phi_{A,K\cup J}\>\right.\\&\left.-n_J(n_K-n_K^2)^{1/2}|\Phi_{A,K}\>+(n_{K\cup J}-n_Kn_J)|\Psi_A\>\right].
\end{align*}
Thus, the action of all $\hat{n}_K$ operators remains in the space $\Ss$ spanned by $\{|\Psi_A\>,|\Phi_{A,J}\>\}$ for $J\in\Jj$ and $A=O,E$. Then, we can calculate the matrix elements for $I,J,K\subset\Jj$
\begin{align*}
\<\Phi_{I,A}|\hat{n}_J|\Phi_{K,A}\>=\frac{\left(n_{I\cup J\cup K}-n_In_{J\cup K}-n_{I\cup J}n_K+n_In_Jn_K\right)}{\sqrt{(n_I-n_I^2)(n_K-n_K^2)}}
\end{align*}
Importantly, these are independent of $A$, and so within the subspace $\Ss$, our number operators act as the block diagonal matrices $\hat{n}_J=\hat{\bar{n}}_J\otimes\Id$. Although the $|\Psi_{J,A}\>$ states are not orthonormal, this fact will not change under the Gram-Schmidt procedure applied equally to both sectors (as the overlap $\<\Psi_{J,A}|\Psi_{K,A}\>$ is again independent of $A$). 

The number operators $\hat{n}_j$ (note small $j$) are mutually commuting, and so they can be simultaneously diagonlised within this subspace $\Ss$. Within this eigenbasis, we can assume that all eigenvectors are separated by eigenvalues of at least one number operator, except that they must come in pairs of opposite parity (due to the $\hat{n}_J=\hat{\bar{n}}_J\otimes\Id$ structure of the number operators). Each of these pairs of eigenvectors define a two-dimensional subspace $\Hh_i$.

The most general form our Majorana operator can take (to commute with each $\hat{n}_J$) is then $\gamma=\sum_{\Hh_i}\gamma_i\sigma_i^\phi$, with $\sigma_i^\phi=\cos(\phi)\sigma_i^z+\sin(\phi)\sigma_i^y$. The coefficients $\gamma_i$ are fixed by transforming back to the original basis (of gram-schmidt orthonormalised $\{|\Psi_A\>,|\Phi_{A,J}\>\}$), and insisting that $\gamma$ contains no terms of the form $|\Psi_A\>\<\Phi_{B,J}|$. In matrix form, the transformation looks as follows:
\begin{equation}
\gamma=\left(\begin{array}{cccc}\gamma_1\sigma^\phi&0&\ldots&0\\0&\gamma_2\sigma^\phi&&0\\\vdots&&\ddots&\vdots\\0&0&\ldots&\gamma_N\sigma^\phi\end{array}\right)\rightarrow\left(\begin{array}{cc}\sigma^\phi&\begin{array}{ccc}0&\ldots&0\end{array}\\\begin{array}{c}0\\\vdots\\0\end{array}&G\end{array}\right).
\end{equation}
Here, $N$ is the dimension of $S$ divided by $2$ (which has a maximum of $2^{|\Jj|}$, but does not necessarily achieve this). Importantly, we see that there are $N$ degrees of freedom in our Majorana, and $N$ conditions fixed by the top row of the original basis. This counting shows that as long as the conditions are not contradictory, at least one solution will exist.

Let $U\otimes\Id$ be the unitary matrix that transforms from the $\hat{n}_j$ eigenbasis to the original basis. Then, our conditions can be written in the form
\begin{equation}
\sum_{j}U^\dag_{1j}\gamma_jU_{jk}=\delta_{1k}.
\end{equation}
If we write $\vec{\gamma}=(\gamma_1\ldots\gamma_N)$, and define the matrix $A_{ij}=U^*_{j1}U_{jk}$, than we have reduced our problem to solving the set of linear equations $A\vec{\gamma}=(1,0,\ldots,0)$. A solution to this will exist as long as the first row of $A$ is not a linear combination of the others. To see that this is always the case, note that the row vectors in $A$ are obtained by multiplying the orthogonal column vectors in $U$ by a diagonal matrix $D_{j,k}=\delta_{j,k}U^*_{j,1}$. This is a shearing operation, which cannot transform orthogonal vectors into linear combinations of each other unless it projects them to the zero vector. And, the first row of $A$ is not projected to the zero vector as it is a termwise square of the elements of the first column vector in $U$, which has norm $1$. Thus, it is always possible to construct at least one localised Majorana operator via the above method when the localisation condition holds.

From the above proof, it is possible to calculate the localised Majorana operator when the localisation condition is explicitly satisfied, by construction of the matrix $A$. In the case of small numbers of sites, we can obtain a closed form of the coefficients of the Majorana operator. Let us write
\begin{equation}
\gamma=\gamma_{\phi}+\sum_{J,K\in\Jj}\left(e^{i\phi}G_{J,K}|\Phi_{O,J}\>\<\Phi_{E,K}|+e^{-i\phi}G_{J,K}|\Phi_{E,K}\>\<\Phi_{O,J}|\right),
\label{eqn:MajoranaConstruction}
\end{equation}
and then solving the linear equations above will give us the coefficients of the $G_{J,K}$. To remove a single site, we have only one coefficient to solve for, and we obtain $G_{1,1}=1$. The solutions for two sites are as follows (with $G_{J,K}=G_{K,J}$)
\begin{align}
A_{1,2}&=\frac{\sqrt{(n_1-n_1^2)(n_2-n_2^2)}}{1-n_1-n_2+n_{12}},\nonumber\\
A_{i,i}&=\frac{(n_i-1)(n_i-n_i^2)}{(n_{12}-n_{\bar{i}})(1-n_1-n_2+n_{12})},\nonumber\\
A_{i,12}&=-\frac{(n_i-1)\sqrt{(n_i-n_i^2)(n_{12}-n_{12}^2)}}{(n_{12}-n_{\bar{i}})(1-n_1-n_2+n_{12})},\nonumber\\
A_{12,12}&=\frac{n_1n_2(1-n_1-n_2+n_{12})+n_{12}(n_1n_2-n_{12})}{(n_{12}-n_2)(1-n_1-n_2+n_{12})}.\nonumber
\end{align}
Here, $i=1$ or $2$, and $\bar{i}$ is the opposite site to $i$. Note that this solution assumes the linear independence of all the $|\Phi_{A,J}\>$; when this is not the case, some of these denominators will be infinite, and a different set of solutions will need to be calculated. These coefficients were used in the construction of the Majorana operators for the double quantum dot.
\end{document}